\documentclass[notitlepage,amsmath,amssymb,twocolumn,pre,tightenlines,nofootinbib,showpacs,superscriptaddress]{revtex4}

\usepackage{amsmath}
\usepackage{amssymb,amsfonts,latexsym}
\usepackage{bm}
\usepackage[mathcal]{euscript}
\usepackage{graphicx}
\usepackage{epsfig}
\usepackage{color}
\usepackage{pifont,wasysym}
\usepackage{textcomp}
\usepackage{comment}

\definecolor{blue}{rgb}{0,0,1}
\definecolor{bleu}{rgb}{0,0,0.8}
\definecolor{bleuf}{rgb}{0,0,0.9}
\definecolor{rougef}{rgb}{0.9,0,0}
\definecolor{green}{rgb}{0,0.5,0}
\definecolor{vert}{rgb}{0,0.8,0}
\definecolor{red}{rgb}{1,0,0}
\definecolor{pink}{rgb}{0.9,0.3,0.7}
\definecolor{azur}{rgb}{0,0.5,0.5}
\definecolor{orange}{rgb}{1,0.5,0.2}
\definecolor{brown}{rgb}{0.5,0,0}

\newcommand{\be}{\begin{equation}}
\newcommand{\ee}{\end{equation}}
\newcommand{\ben}{\begin{equation*}}
\newcommand{\een}{\end{equation*}}
\newcommand{\ba}{\begin{eqnarray}}
\newcommand{\ea}{\end{eqnarray}}

\let\grad\nabla
\newcommand{\Pe}{\mbox{Pe}}
\newcommand{\Ub}{\mathbf{U}}
\newcommand{\mean}[1]{\left\langle #1\right\rangle}
\newcommand{\pard}[2]{\frac{\partial #1}{\partial #2}}
\newcommand{\eb}{\mathbf{e}}

\begin{document}
\graphicspath{{./Figures/}}

\title{Self-propulsion of pure water droplets by spontaneous Marangoni stress driven motion}
\author{Ziane Izri}
\author{Marjolein N. van der Linden}
\affiliation{EC2M, UMR Gulliver 7083 CNRS, ESPCI ParisTech, PSL Research University, 10 rue Vauquelin, 75005 Paris, France}
\author{S\'ebastien Michelin}
\affiliation{LadHyX -- D\'epartement de M\'ecanique, Ecole Polytechnique -- CNRS, 91128 Palaiseau, France}
\author{Olivier Dauchot}
\email[]{olivier.dauchot@espci.fr}
\affiliation{EC2M, UMR Gulliver 7083 CNRS, ESPCI ParisTech, PSL Research University, 10 rue Vauquelin, 75005 Paris, France}

\begin{abstract}
We report spontaneous motion in a fully bio-compatible system consisting of pure water droplets in an oil-surfactant medium of squalane and monoolein. Water from the droplet is solubilized by the reverse micellar solution, creating a concentration gradient of swollen reverse micelles around each droplet. The  strong advection and weak diffusion conditions allow for the first experimental realization of spontaneous motion in a system of isotropic particles at sufficiently large P\'eclet number according to a straightforward generalization of a recently proposed mechanism ~\cite{Michelin:2013gv,Michelin:2014hv} 
Experiments with a highly concentrated solution of salt instead of water, and tetradecane instead of squalane, confirm the above mechanism. The present swimming droplets are able to carry external bodies such as large colloids, salt crystals, and even cells.
\end{abstract}


\maketitle

--- The recent surge of interest in active systems has driven an intense research effort towards the design of self-propelled polar particles, including walking grains~\cite{Deseigne:2012kn,Deseigne:2010gc}, rolling~\cite{Bricard:2013jq} or skating~\cite{Palacci:2013eu} colloids, and a variety of swimmers~\cite{Paxton:2006dm,poesio2009,Palacci:2010hk,Theurkauff:2012ui,Poon:2013vz}. 
The latter often take the form of Janus-like colloids~\cite{Walther:2008en}, named after the two-faced Roman god, because their motion originates from the asymmetry of their surface properties~\cite{Howse:2007ed,Theurkauff:2012ui,Palacci:2010hk}. 
An alternative design of artificial swimmers consists of active droplets, either on interfaces~\cite{Chen:2009kb,Yabunaka:2012vk} or in bulk fluid~\cite{Toyota:2009cg,hanczyc2007fatty,Thutupalli:2011bv,Herminghaus:2014cg}. Droplets are particularly interesting systems since they are extensively used in microfluidic devices as (bio-)chemical reactors~\cite{deMello:2006bq,Baraban:2011id}. Replacing the external flow transport of the droplets by self-propulsion would open new ways towards yet unexplored applications.

The self-propulsion mechanism of swimming droplets~\cite{Toyota:2009cg,hanczyc2007fatty,Thutupalli:2011bv,Herminghaus:2014cg} has its origin in the Marangoni flow induced by a surface-tension gradient. In most cases, this gradient is maintained through specific chemical reactions, including the hydrolysis~\cite{Toyota:2009cg,hanczyc2007fatty} or the bromination of the surfactant~\cite{Thutupalli:2011bv}. Liquid crystal droplets, stabilized by ionic surfactant, were also shown to develop spontaneous motion under certain circumstances of adsorption-depletion of the surfactant at the droplet interface~\cite{Herminghaus:2014cg}. Apart from being specific, these conditions may also be undesirable due to possible interactions between the chemicals and the products that are to be transported in the droplets.

Recently, it was shown theoretically that at sufficiently large P\'eclet number (strong advection, weak diffusion) the nonlinear interplay between surface osmotic flows and solute advection can produce spontaneous and self-sustained motion of isotropic particles~\cite{Michelin:2013gv,Michelin:2014hv}. In principle, sufficiently large droplets generating a solute of sufficiently large molecules or nanoparticles should thus self-propel without requiring any sort of chemical reaction.

\begin{figure}[t!] 
\center
\begin{minipage}[c]{.55\linewidth}
	\includegraphics[width=\columnwidth]{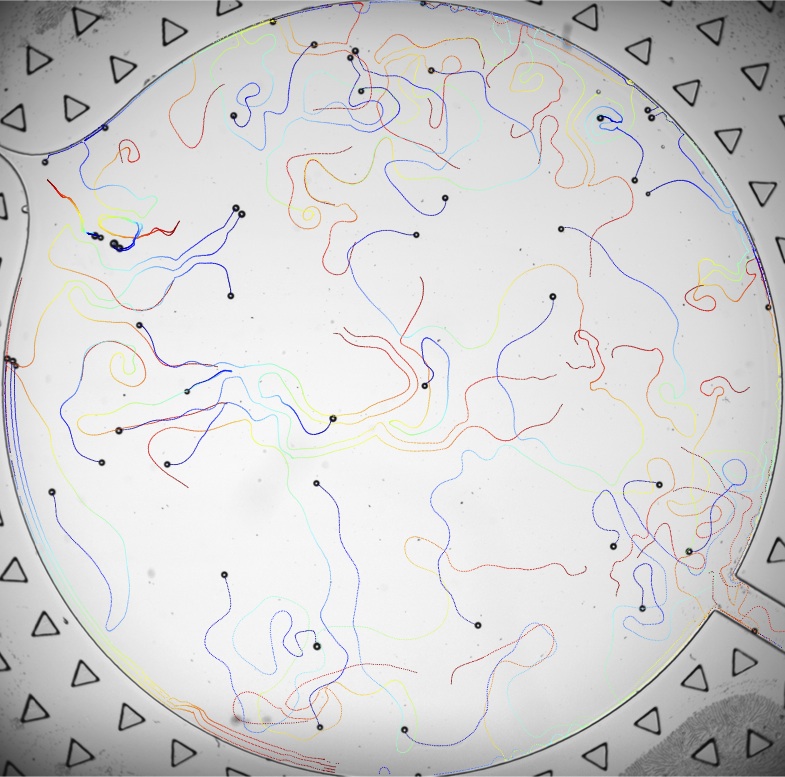}
\end{minipage}
\begin{minipage}[c]{.42\linewidth}
	\includegraphics[width=0.48\columnwidth]{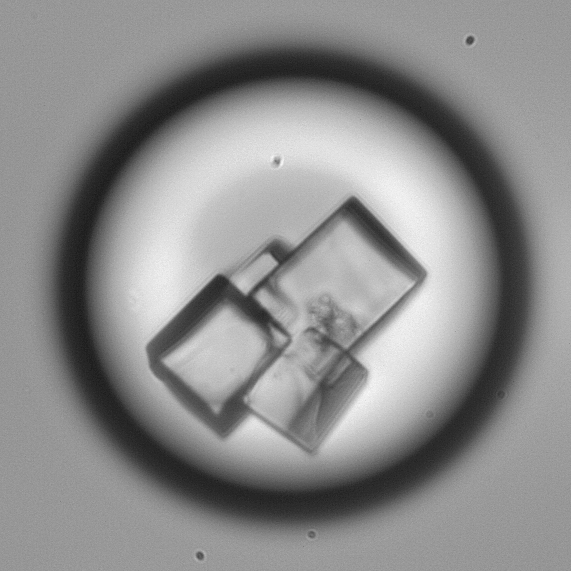}
	\includegraphics[width=0.48\columnwidth]{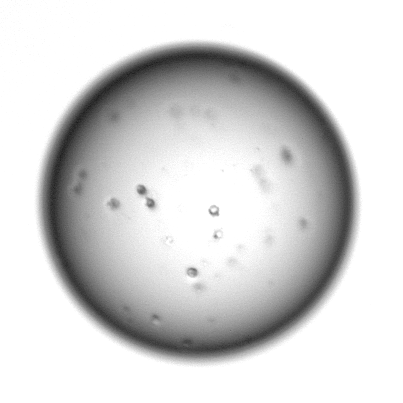}
	\includegraphics[width=\columnwidth]{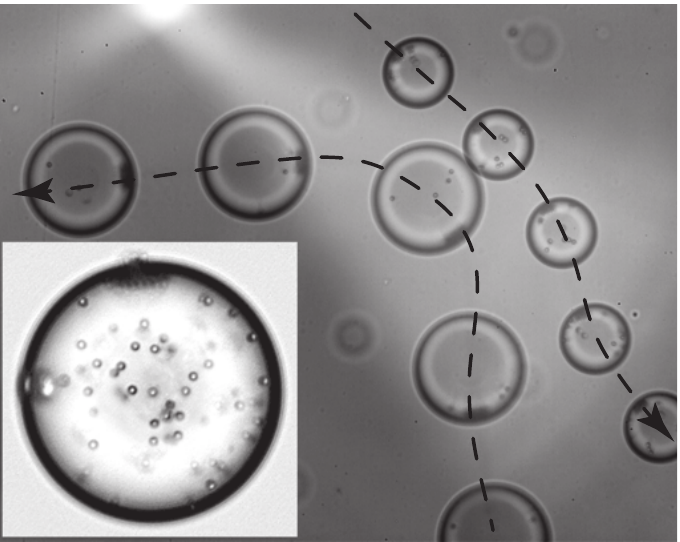}\\
\end{minipage}
\begin{flushleft}
\hspace{0.3\columnwidth}(a)\hspace{0.45\columnwidth}(b)
\end{flushleft}
\vspace{-0.2cm}
\caption{(a) {\bf Pure water droplet motion:} Trajectories of $\simeq 50$ water droplets in the observation room (diameter 1 cm) filled with Sq./25 mM MO, recorded during 500 seconds. The droplet trajectory is color-coded with the time preceding its present location (color online) (see movie in Supp.\ Mat.\ \cite{suppmat}). (b) {\bf Swimming droplets as microfluidic carriers:} Transport of salt crystals (top-left), Dami cells~\cite{greenberg1988} (top-right) and colloids (bottom).}
\vspace{-0.5cm}
\label{fig:setup} 
\end{figure}

In this Letter, we demonstrate experimentally the self-propulsion of \emph{pure water} droplets in a bio-compatible oil (squalane) - surfactant (monoolein) medium (fig.\ 1(a)). To the best of our knowledge, our system also constitutes the simplest realization of spontaneous motion in a system of isotropic particles as predicted in~\cite{Michelin:2013gv,Michelin:2014hv}. Replacing water with a saturated solution of salt, or squalane with tetradecane, we prove the robustness of the swimming mechanism.  Finally, we take advantage of this robustness to illustrate the carrier function of these new swimming droplets, by transporting large colloids, salt crystals (which form in the saturated salt solution) and Dami cells~\cite{greenberg1988} (fig.\ 1(b)) inside the droplets.

--- The experimental system consists of pure water droplets (milli-Q) with typical radii $a$ of 20--60 \textmu m in a continuous oil-surfactant phase consisting of a solution of 25 mM of the nonionic surfactant monoolein (MO; 1-oleoyl-\textit{rac}-glycerol, 99\%, Sigma) in the oil squalane (Sq; 99\%, Aldrich). The surfactant concentration is far above the critical micellar concentration (cmc) for MO in Sq ($1.5$ mM \cite{thutupalli2011}). We also use droplets consisting of an almost saturated solution of 26 wt\% NaCl (Sigma-Aldrich) in water and replace the continuous phase with tetradecane (Td; $\ge99$\%, Aldrich) with 25 mM MO. 
The microfluidic device is made up of a 150-\textmu m layer of UV-curing glue (Norland Optical Adhesive No.\ 81) on top of a cover slip held between two microscope slides~\cite{bartolo2008}. The device comprises a T-junction for droplet production with channel section of $50 \times 50\, \mu m^2$ and a circular observation chamber with a diameter of 1.0 cm and a height of 150 \textmu m. Initially, the whole system is filled with the oil/surfactant solution. Water droplets are then produced at the T-junction using typical flow rates of 10--50 \textmu L/h and sent to a trash channel until the desired droplet size and density are obtained. 
We then redirect the flow towards the observation chamber and send 10--100 droplets into the chamber by greatly increasing the oil flow rate (to typically 1000 \textmu L/h). When the droplets reach the center of the chamber we stop all flows. Images are recorded (Falcon II camera, Teledyne Dalsa) on a Nikon AZ100 macroscope (AZ Plan APO 1$\times$ NA 0.1 objective) for 2 hours at $3\times$ magnification and an acquisition rate of 1 frame per second. 
We obtain the droplet coordinates by processing the images in Labview using object and circle detection algorithms. Droplet trajectories are tracked using a Matlab algorithm adapted from~\cite{crocker1996}.

\begin{figure}[t!] 
\center
\includegraphics[width=0.49\columnwidth,height=0.49\columnwidth]{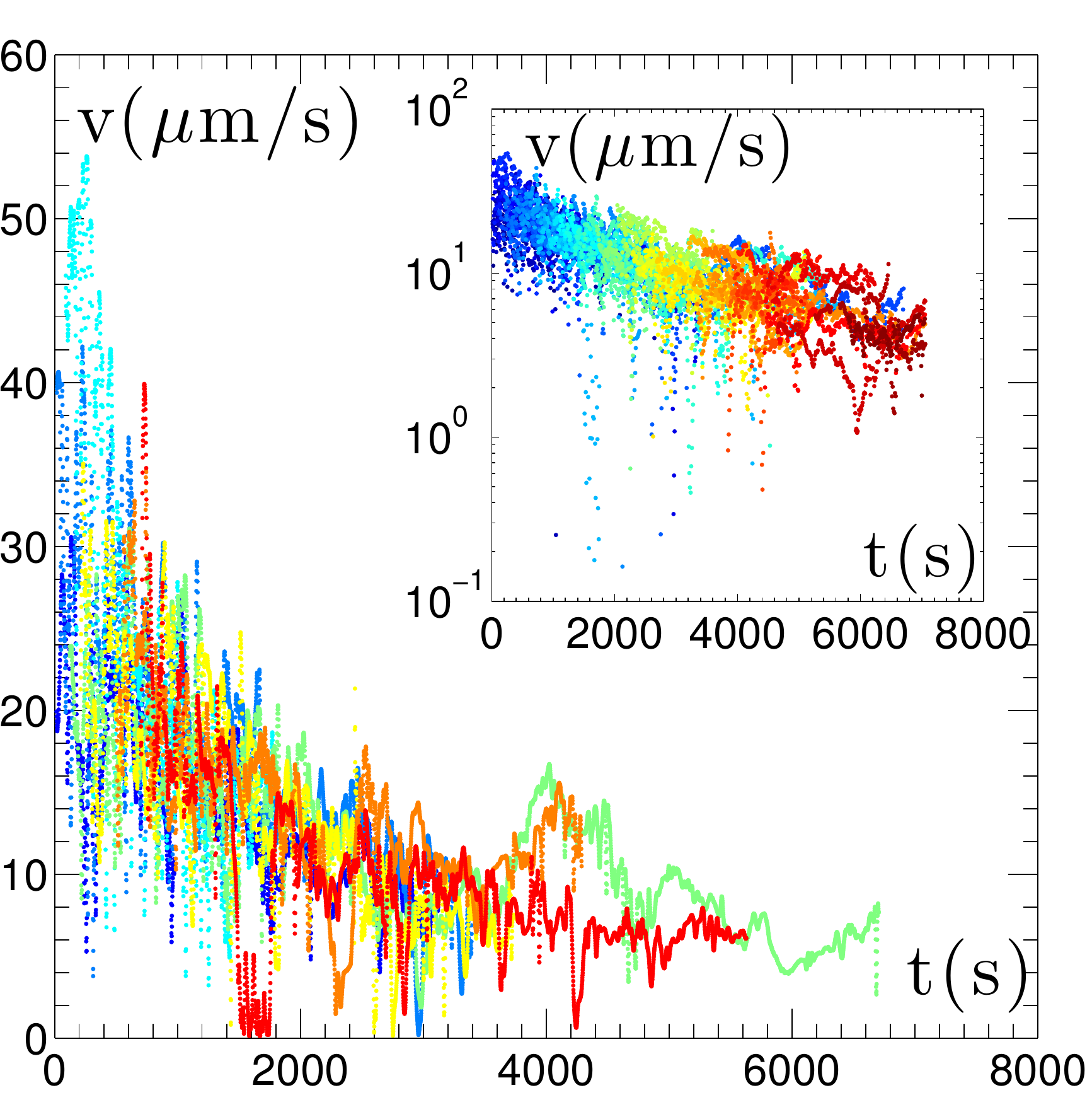}
\includegraphics[width=0.49\columnwidth,height=0.49\columnwidth]{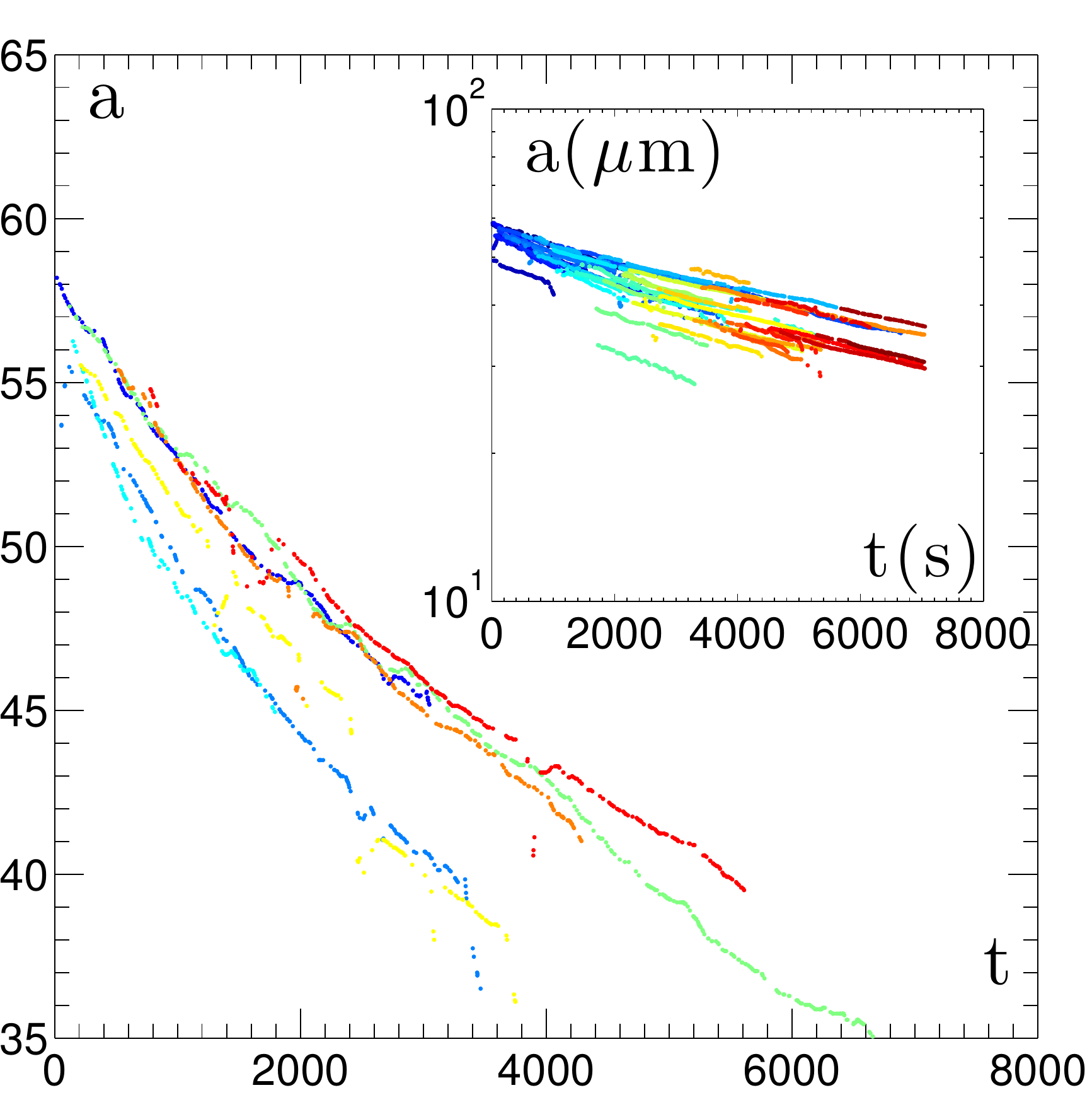}
\vspace{-0.5cm}
\begin{flushleft}
\hspace{0.22\columnwidth}(a)\hspace{0.45\columnwidth}(b)
\end{flushleft}
\vspace{-0.5cm}
\caption{{\bf Pure water droplet motion~:} (a) Velocity and (b) diameter versus time for a selection of 8 trajectories. Insets: lin-log plot for a selection of 35 trajectory parts. (color online)}
\vspace{-0.3cm}
\label{fig:swim} 
\end{figure}

---  First we describe the droplet dynamics in the Sq/MO solution. After the fluid flows are stopped, the droplets move spontaneously in random directions. Fig.~\ref{fig:setup}(a) displays a picture of 50 droplets in the observation chamber, together with their trajectories recorded during a period of 500 seconds before the picture is taken. The droplets exhibit curved trajectories with a typical persistence length of the order of 500 \textmu m. Interactions between the droplets are rather involved: we observe repulsion when the droplets move fast, but also attraction when they are slower. Some droplets form pairs and swim in parallel, (see center of fig.~\ref{fig:setup}-a)). In the following we concentrate on dilute systems and the short-time dynamics. The characterization of the long-time dynamics and possible collective effects are left for future work.

Typically, initial velocities are in the range 10--50 \textmu m/s and the swimming motion last for 2 hours, during which the velocity decays exponentially with time, with a characteristic decay time $\tau_v \simeq  3500$ s (fig.~\ref{fig:swim}(a)). As a result the trajectory length extends up to several thousands droplet diameters, a ``cruising range'' never achieved before. A remarkable observation is that the droplet size also decreases as a function of time (fig.~\ref{fig:swim}(b)). The exponential decay is not as clear as for the velocity, but we can still estimate a characteristic decay time $\tau_a \simeq 8000$ s. From this very basic observation, we infer that there is, in one form or another, a net flux of water coming from the droplet at an almost constant rate $\kappa = a(0)/\tau_a \simeq 5. 10^{-3} $ \textmu m s$^{-1}$, where $a(0)$ is the initial droplet radius. A natural question is how the velocity scales with the droplet radius. However, as can readily be seen in fig.~\ref{fig:swim}(a), there are very strong fluctuations of the velocity : a droplet may slow down by more than a factor of 100 before recovering its nominal velocity. 

\begin{figure}[t!] 
\center
\includegraphics[width=0.49\columnwidth,height=0.52\columnwidth]{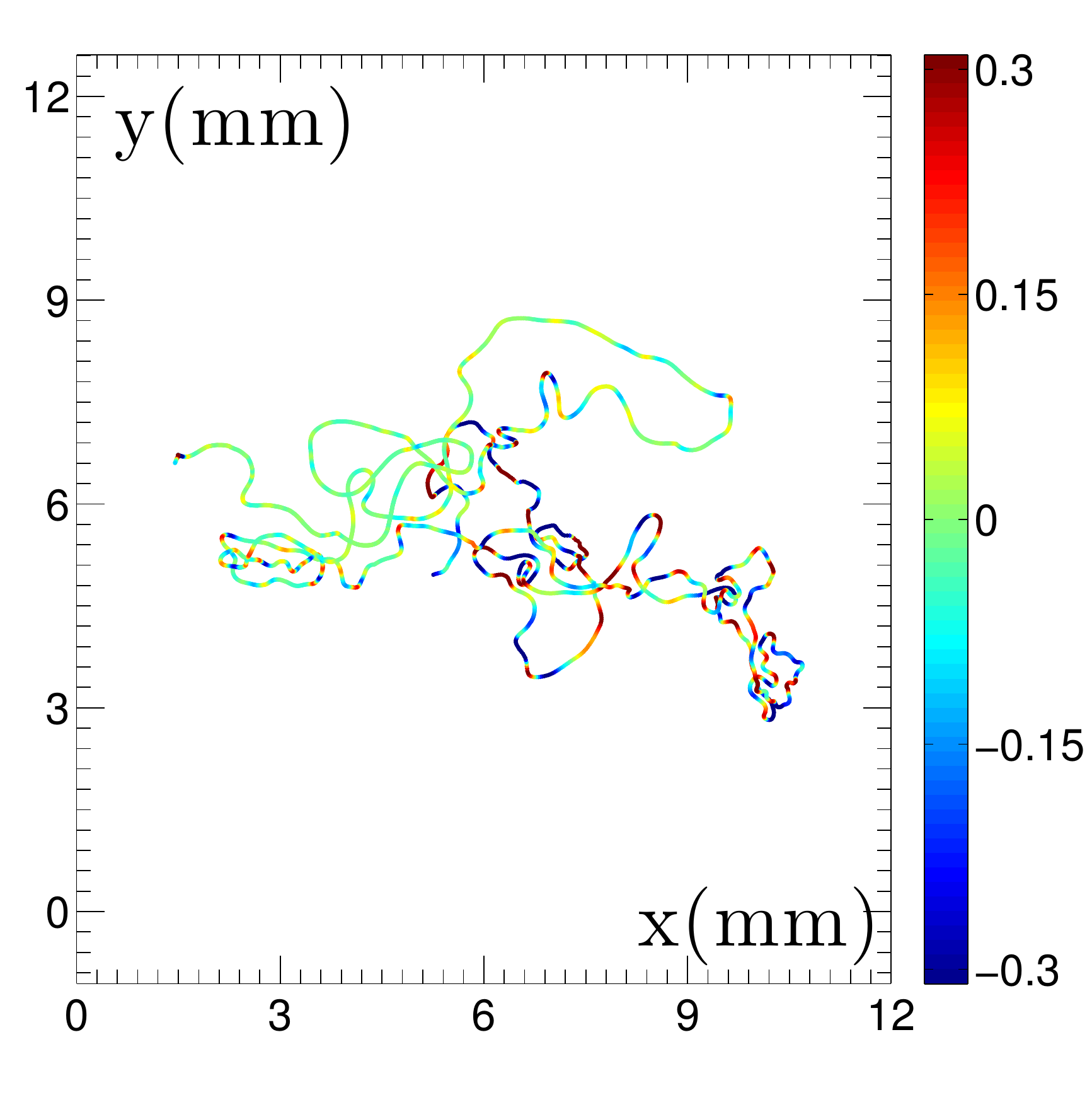}
\includegraphics[width=0.49\columnwidth,height=0.52\columnwidth]{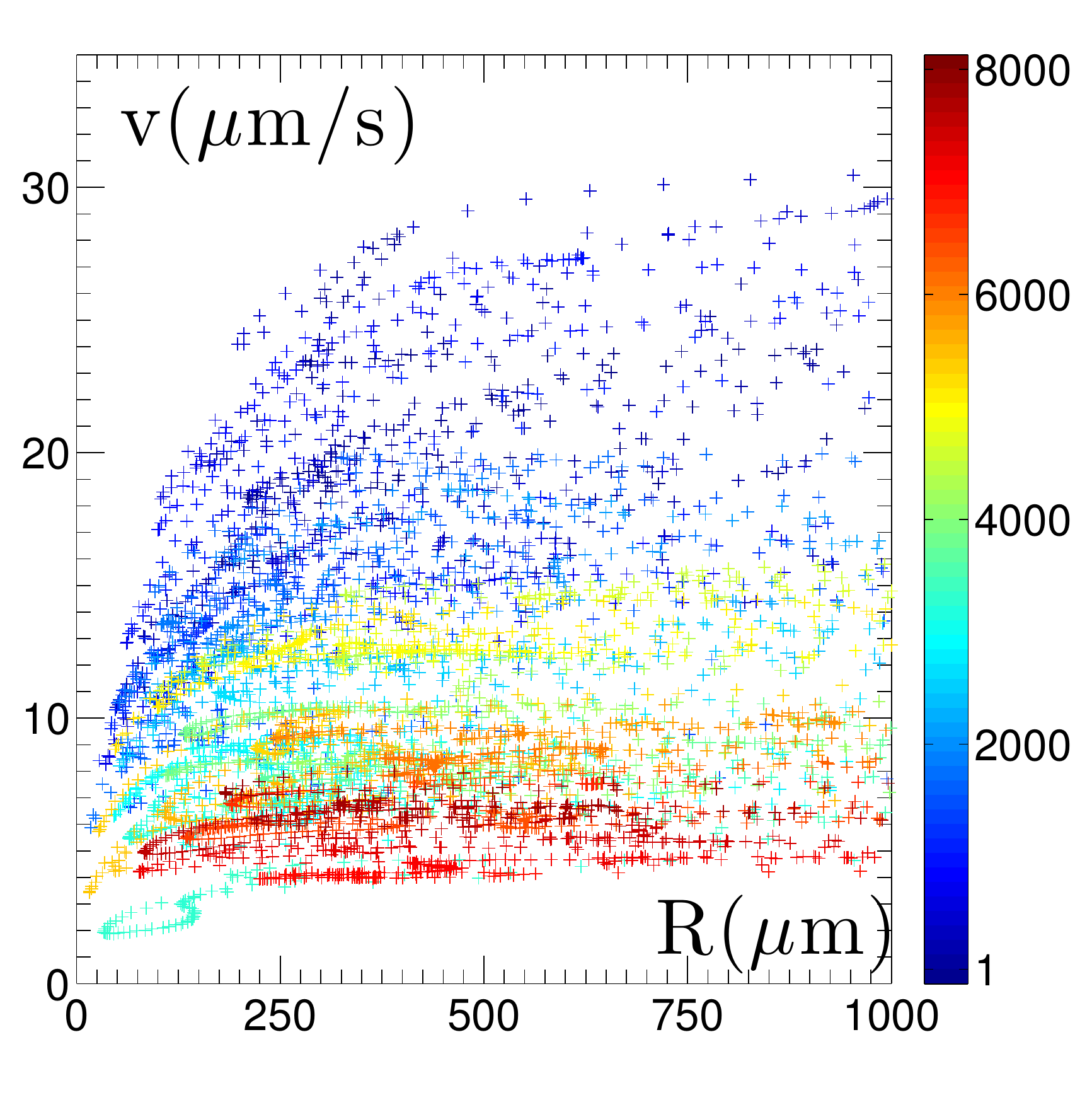}
\vspace{-0.7cm}
\begin{flushleft}
\hspace{0.2\columnwidth}(a)\hspace{0.45\columnwidth}(b)
\end{flushleft}
\vspace{-0.5cm}
\caption{{\bf Correlation between velocity and local radius of curvature of the trajectory~:} (a) A typical trajectory colored by the instantaneous tangential acceleration (\textmu m s$^{-2}$). (b) Velocity as a function of the radius of curvature of the trajectory shown in (a). The color code indicates time (s) from the beginning of the trajectory. (color online)}
\vspace{-0.3cm}
\label{fig:V_vs_Curv} 
\end{figure}

Such fluctuations are completely absent from the radius dynamics (fig.~\ref{fig:swim}(b)), suggesting the existence of at least one other parameter controlling the instantaneous droplet velocity.
Figure~\ref{fig:V_vs_Curv}(a) represents a typical trajectory, which has been colored according to the instantaneous tangential acceleration of the droplet. Straight parts of the trajectories have small tangential acceleration, while curved parts of the trajectory are preceded by a negative acceleration and followed by a positive one. This is confirmed in fig.~\ref{fig:V_vs_Curv}(b), where the instantaneous velocity $v$ is plotted as a function of the instantaneous radius of curvature $R$ (the time is color-coded from blue to red). Each time the radius of curvature decreases below say $500$ \textmu m, the velocity decreases strongly and subsequently increases again with increasing radius of curvature. For a radius of curvature larger than $500$ \textmu m the velocity depends only on time. We checked that after filtering the velocity data and retaining only the parts of the trajectories with $R>500$ \textmu m, the fluctuations observed in fig.~\ref{fig:swim}(a) are suppressed. Note that, in the absence of inertia, one should not interpret these observations in terms of ``cautious driving''. Here, the curvature presumably results from the repulsion between the droplets, which in addition slows the droplets down when they approach and speeds them up when they separate.

\begin{table}
\begin{center}
\begin{tabular}{|c|c|c|c|}
Discrete phase & Continuous phase & Surfactant & Motion \\
\hline 
H$_2$O	& 	Sq 		& 	MO		& 		yes 	\\
H$_2$O	& 	Sq 		& 	MO ($<$ cmc) 	& 		no 	\\
H$_2$O 	& 	Sq		& 	Span 80			& 	no   	\\
H$_2$O 	& 	Sq		& 	oleic acid			& 	no   	\\
\hline
H$_2$O	& 	Td 		& 	MO    	& 	yes	\\
H$_2$O		& 	 water-saturated 	& 	MO	&	no	\\
		&	Sq 		&  			&	 	\\
\hline
H$_2$O + salt	& 	Sq 				& 	MO	& 	yes \\	
H$_2$O + malonic acid 	& 	Sq 		& 	MO	& 	yes \\	
H$_2$O + inhibited BZ	& 	Sq 		& 	MO	& 	yes \\
\hline
\end{tabular}
\end{center}
\caption{Realization of swimming motion for various water-oil/surfactant systems. The surfactant concentration is above the cmc, unless indicated otherwise. (see text for details)}
\label{tab:systems}
\vspace{-0.5cm}
\end{table}

\begin{figure}
\center
\includegraphics[width=0.49\columnwidth]{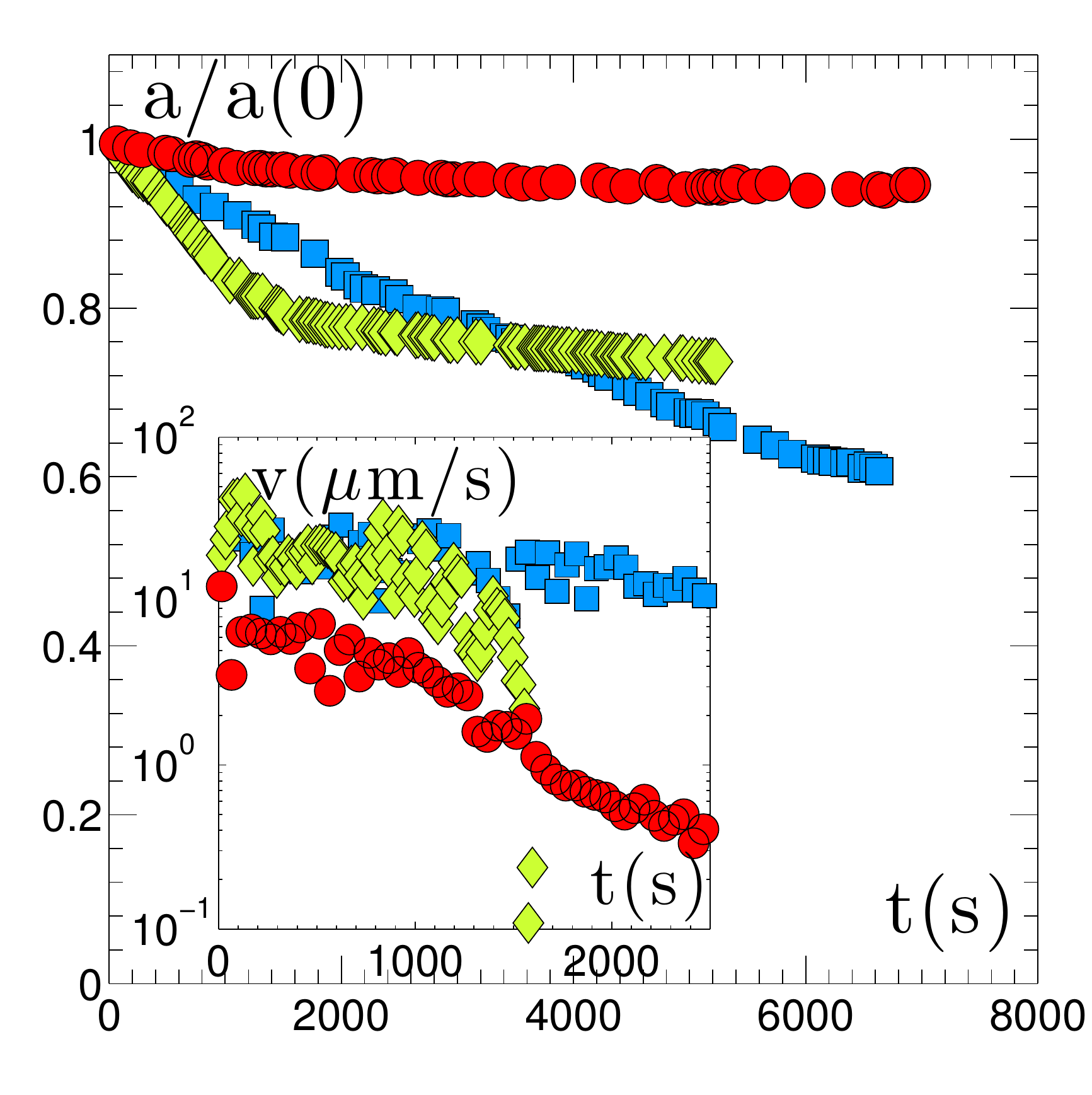}
\includegraphics[width=0.49\columnwidth]{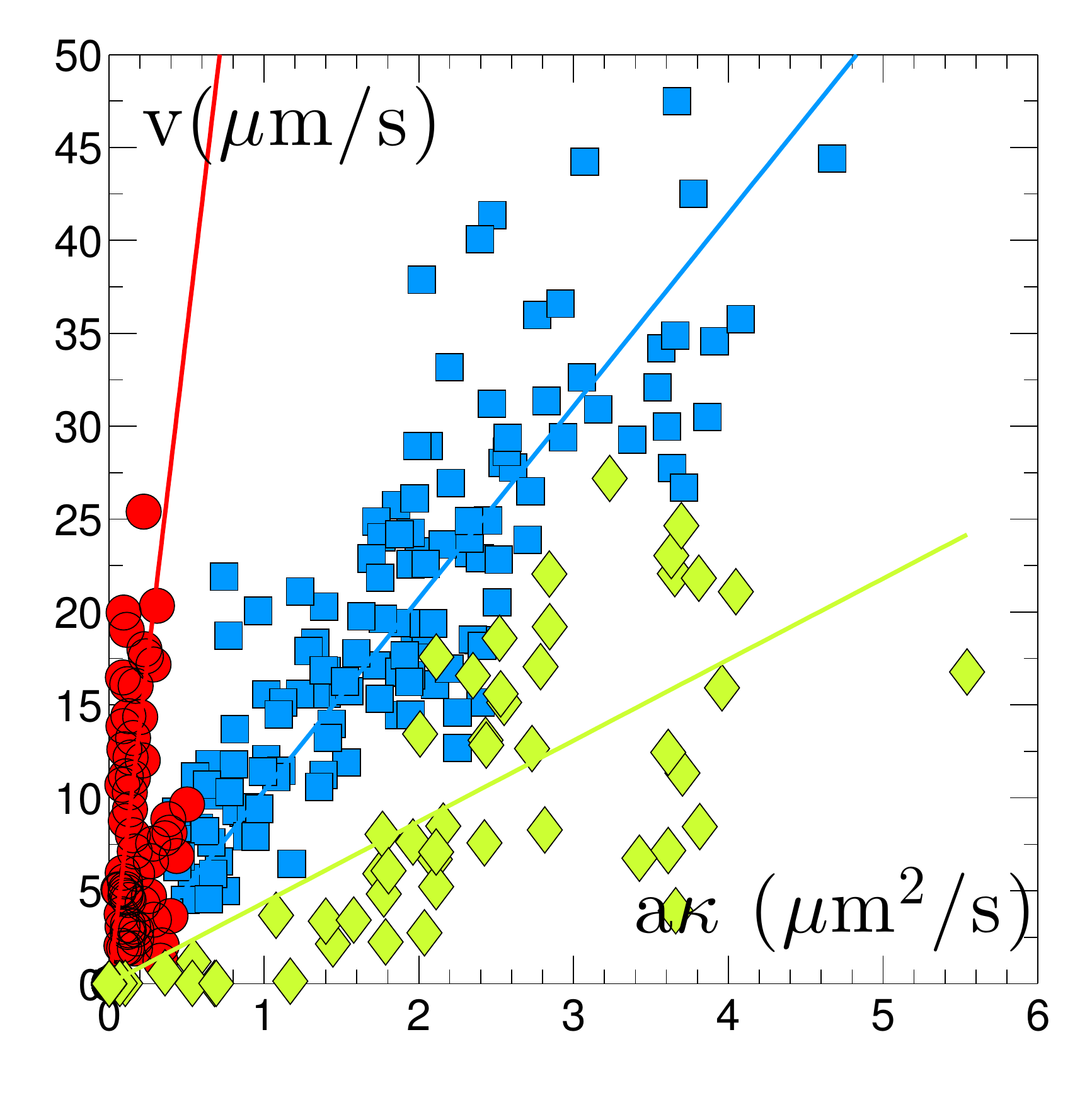}
\vspace{-0.7cm}
\begin{flushleft}
\hspace{0.24\columnwidth}(a)\hspace{0.45\columnwidth}(b)
\end{flushleft}
\vspace{-0.5cm}
 \caption{{\bf Comparison of three different systems~:} (a) Decay of the relative droplet radius versus time for three droplets under different conditions: water in Sq/MO (blue $\square$), water in Td/MO (green $\diamond$), water/26 wt$\%$ NaCl in Sq/MO (red $\circ$). Inset: velocity versus time for the same three droplets. (b) Parametric plot of the velocity versus $a\times \kappa$ for droplets under the same three different sets of conditions; each point represents a different droplet during a different period of time corresponding to a straight part of its trajectory.}
\vspace{0.0cm}
\label{fig:systems}
\end{figure}

Before discussing the mechanism of self-propulsion, we consider the robustness of the phenomena. Table~\ref{tab:systems} lists the various systems we have examined. We have separately varied the surfactant, the oil and the composition inside the droplet. The present list is by no means exhaustive, but aims at testing basic variations in search of the essential ingredients of the underlying swimming mechanism. The first result is that MO as a surfactant is a key ingredient of the microscopic mechanism responsible for swimming. Neither Span 80, nor oleic acid, which have the same apolar tail, but different polar head groups, leads to swimming motion \cite{surf}. Furthermore, it is crucial that the MO concentration is above the cmc, telling us that micelles play a key role in the physico-chemical mechanism. When Sq is replaced with Td, the swimming motion still occurs, suggesting that the choice of the continuous phase is not as crucial. 
However, the use of water-saturated Sq, obtained by keeping the Sq/MO in contact with water during several days, suppresses the swimming motion. This clearly indicates that gradients of water (in some form) around the droplet are essential to the swimming mechanism. Coming now to the discrete phase, we added NaCl to the water in order to test whether osmotic pressure, which tends to keep water inside the droplet, would prevent swimming. As we show below, the presence of salt alters the swimming motion, but does not suppress it, even at a salt concentration of 26 wt\%, close to saturation. Finally, inspired by the work of Thutupalli \textit{et~al.\ }on swimming water droplets~\cite{Thutupalli:2011bv}, we added successively all the compounds of the Belousov-Zhabotinski (BZ) reaction (sulfuric acid, sodium bromate, malonic acid) to the water, \emph{except} for the catalyst, using the same concentrations as in~\cite{Thutupalli:2011bv}. Amazingly, the droplets are still swimming, suggesting that the mechanism at play in the present work may also be present in the system of~\cite{Thutupalli:2011bv}. Focusing more quantitatively on the most relevant systems described above, namely the [water in Sq/MO], the [water in Td/MO] and the [salt-saturated water in Sq/MO] systems, it is observed (see fig.~\ref{fig:systems}(a) and table~\ref{tab:comp}) that a larger decay rate of the droplet size $\left<\kappa\right>$ corresponds to a faster initial swimming velocity $\left<v(0)\right>$, and a shorter duration of the swimming motion $T_{\rm{swim}}$.

\begin{table}
\vspace{-2mm}
\begin{center}
\begin{tabular}{|c|c|c|c|c|}
System & $\left<D(0)\right>$ & $\left<\kappa(0)\right>$ (\textmu m/s) & $\left<v(0)\right>$ (\textmu m/s) & $T_{\rm{swim}}$ \\
\hline 
(1) &	 $91\pm 5$ & $4.5\,10^{-2}\pm 5\,10^{-3}$ & $20 \pm 2$ & $2 \rm{h}$ \\ 
(2) & $93\pm 5$ & $7.5\,10^{-2}\pm 5\,10^{-3}$ & $22 \pm 2$ & $1/2 \rm{h}$ \\
(3) & $63\pm 5$ & $1.2\,10^{-2}\pm 5\,10^{-3}$ & $11 \pm 2$ & $> 2 \rm{h}$ \\  
\hline
\end{tabular}
\end{center}
\vspace{-3mm}
\caption{Swimming motion characteristics at initial times for three systems of interest: (1) H$_2$O in Sq/MO, (2) H$_2$O in Td/MO, (3) H$_2$O + 26 wt\% NaCl. (see text for details)}
\label{tab:comp}
\vspace{-0.5cm}
\end{table}

--- From the macroscopic observations (the droplet radius decreases in time) we know that, in some form or another, water leaves the droplets, hence producing a gradient of �solute� outside each droplet. This is a situation very similar to the one investigated theoretically in~\cite{Michelin:2013gv,Michelin:2014hv}: a spherical particle of radius $a$ emits ($A > 0$) or captures ($A < 0$) a solute with a uniform surface emission rate (activity) $A$. The solute interacts with the spherical particle on a small length scale $\lambda\ll a$, giving rise, whenever a surface gradient of solute $\grad_\parallel C$ develops, to a slip velocity $v^s = M\grad_\parallel C$ that drives a net flow outside the particle. The phoretic mobility $M \approx \pm k_B T \lambda^2/ 2\eta_o$, where $\eta_o$ is the viscosity of the surrounding fluid, can either be positive or negative depending on the particle surface-solute interaction potential~\cite{Anderson:1989vq}. The trivial solution to the coupled Stokes flow and advection-diffusion of the solute corresponds to an isotropic solute concentration and no fluid motion. However, when $AM>0$, this isotropic solution is linearly unstable above a critical P\'eclet number $\Pe = U^*a/D$ where $U^* = |AM|/D$ is the characteristic auto-phoretic velocity, and $D$ the diffusion coefficient of the solute. This leads to a spontaneous symmetry breaking of the concentration field and propulsion (see fig. 5, with $\eta_i\rightarrow\infty$).

\begin{figure}
\center
\includegraphics[width=0.98\columnwidth]{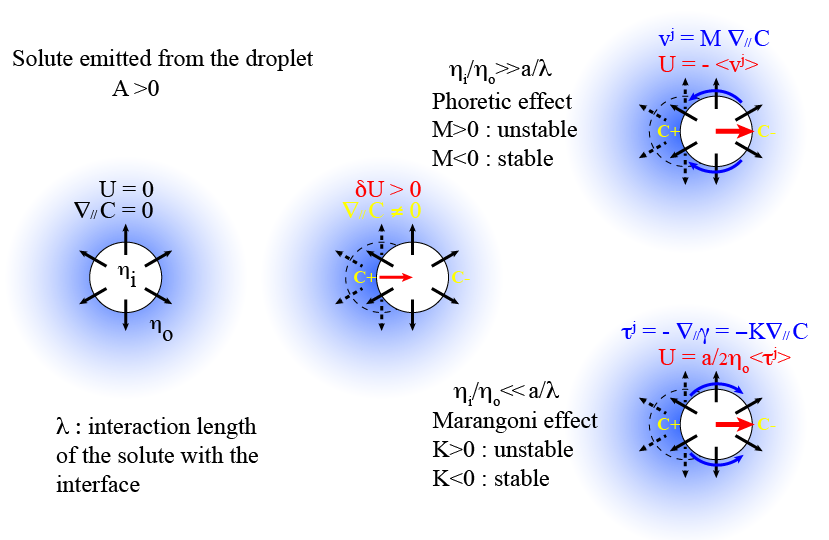}
\vspace{-0.3cm}
 \caption{\label{fig:mech} {\bf Swimming mechanism } behind the spontaneous auto-phoretic  and Marangoni-driven motions of an isotropic droplet.}
\vspace{-0.5cm}
 \end{figure}

Extending this result to active droplets follows easily from two essential properties shared with the autophoretic particle problem: (i) the solute is released at the droplet's surface and advected by the outer flow, and (ii) this flow results from tangential motion of the surface of an impermeable force-free spherical body. The two configurations differ however on the existence of a flow within the droplet with viscosity $\eta_i$, and on the hydrodynamic boundary conditions at the interface: the slip velocity is replaced by an analogous velocity jump $v^j = M\grad_\parallel C$ (phoretic effect) complemented by a tangential stress jump $\tau^j =-\grad_\parallel\gamma$ across the interface (Marangoni effect). The surface tension gradient originates from a number of physico-chemical mechanisms, including the solute-surface interaction or potentially the desorption of the surfactant from the interface~\cite{Herminghaus:2014cg}. The common cause for all these mechanisms is $\grad_\parallel C$ and to linear order, we expect $\grad_\parallel \gamma=K\grad_\parallel C$, where $K \approx \pm k_BT \lambda$. Under this assumption, the droplet velocity is now obtained as~\cite{Anderson:1989vq}
\begin{equation}
\label{eq:velocity}
\Ub=-M'\mean{\grad_\parallel C},\quad \textrm{with   }M'\equiv\frac{aK+3\eta_i M}{2\eta_o+3\eta_i}.
\end{equation}
where $\langle \cdot\rangle$ denotes the average over the  interface.

Repeating the analysis of~\cite{Michelin:2013gv}, we consider an axisymmetric perturbation of the isotropic concentration distribution $\bar{C}(r)=Aa^2/Dr$. Its first azimuthal moment, $C_1'(r,t)$ satisfies at leading order 
\begin{align}
\label{eq:perturbation}
\pard{C_1'}{t}-\frac{D}{r^2}\left[\pard{}{r}\left(r^2\pard{C_1'}{r}\right)-2C_1'\right]&=\frac{Aa^2U}{Dr^2}\left(\frac{a^3}{r^3}-1\right).
\end{align}
Because $\mean{\grad_\parallel C}=2C_1'(a,t)\eb_z/3a$, Eqs.~\eqref{eq:velocity}--\eqref{eq:perturbation} are strictly identical to the dimensional form of Eqs.~(10)--(11) of~\cite{Michelin:2013gv}. The linear stability results are therefore directly applicable here provided the phoretic mobility $M$ is replaced by $M'$. The change in boundary conditions impacts however the nonlinear dynamics and steady-state velocity (see Supplemental Material \cite{suppmat}).

The relative importance of Marangoni and phoretic effects is given by the comparison of the viscosity ratio $\eta_i/\eta_o$ with the length scale ratio $a/\lambda$. For droplets in general, and in the present case in particular ($\eta_i/\eta_o=1/36$), Marangoni effects largely dominate so that $M'=\frac{aK}{2\eta_o}$. The spontaneous propulsion of the droplets indicates that the delicate balance of the physico-chemical mechanisms at play ensures $K>0$, the necessary condition for the linear instability to take place, provided the P\'eclet number $\Pe=\frac{AM'a}{D^2}$ is greater than $\Pe_c=4$~\cite{Michelin:2013gv}.

In order to proceed, we specify in what form water leaves the droplets. We have seen that empty reverse micelles of MO are necessary for the realization of the swimming motion. Following~\cite{shinoda1967, Herminghaus:2014cg}, we propose that the water is solubilized by the reverse micelles that are present in the continuous oil phase, forming swollen reverse micelles that act as the `solute'. This is consistent with the observation that using a water-saturated Sq/MO oil phase prevents both the shrinkage and the swimming of the droplets. 
Furthermore, dynamic light scattering experiments on the Td/MO oil phase that had been in contact with water droplets for 1 to 3 hours, which were removed by centrifugation, reveal a typical radius $\delta\approx10$ nm for the swollen reverse micelles, while no objects of this size are present in the native oil phase.

In the present context, the activity $A$ is easily related to the decrease rate of the droplet radius $\kappa=|\frac{\mathrm{d}a}{\mathrm{d}t}|$ by equating  the number of swollen reverse micelles of radius $\delta$ formed per unit time, $\frac{\mathrm{d}N}{\mathrm{d}t} = 4 \pi a^2 A$, to the change in volume of the droplet divided by the volume of a swollen reverse micelle $ \frac{1}{\delta^3} \frac{\mathrm{d}a^3}{\mathrm{d}t} = 3 \kappa \frac{a^2}{\delta^3} $, from which it follows that $A=\frac{3}{4\pi}\frac{\kappa}{\delta^3}$. Taking the diffusion coefficient of the solute $D=k_\mathrm{B}T/(6\pi\eta_\mathrm{o}\delta)$, we then obtain for the typical droplet velocity:
\begin{equation}
\label{eq:Ustar2}
U^*=\frac{AM'}{D} \approx \frac{9}{4} \kappa \frac{a \lambda}{\delta^2}.
\end{equation}
\noindent
The linear scaling of the droplet velocity with $\kappa a$ is indeed observed in fig.~\ref{fig:systems}(b). Furthermore, assuming $\lambda\simeq\delta$ and $\delta\approx10$ nm, we find a characteristic velocity of a few tens of $\mu$m/s and $\mathrm{Pe}>>1 $, in agreement with the observations and the linear instability condition.

---  In summary, we have established the first experimental evidence of spontaneous swimming of pure water droplets. The conditions of swimming are threefold: (i) water droplets must be stabilized in an oil medium with surfactant above CMC; (ii) the surfactant inverse micelles must be prone to extract water from the droplets; (iii) the P\'eclet number must be large enough (large droplets, high oil viscosity, fast kinetics of the water transfer to the micelles). Apart from confirming a very general instability mechanism, it opens new ways to a plethora of applications. As a first step in this direction, we have demonstrated the versatility of these droplets as universal carriers: fig.~\ref{fig:setup}(b) illustrates the transport of colloids, salt crystals and cells. We found that the swimming of the droplets was also maintained for pH $\in [3-11]$. Such robustness indicates that there is room for optimization, which in turn calls for a detailed investigation of the physico-chemical mechanisms at play.


{\em Acknowledgements} --- 
The authors thank Denis Bartolo for his help with microfluidics and enlightening discussions, the MMN lab for their help in designing microfluidic devices, B\'{e}reng\`{e}re Abou for her support in performing the DLS measurements and Antoine Blin for providing the cells. The project was supported by ANR MiTra, and MNvdL was sponsored by a postdoctoral fellowship from DIM ISC. 

ZI and MNvdL contributed equally to this work.


\begin{thebibliography}{35}
\bibitem[{sur()}]{surf}
\bibinfo{note}{The polar head group affects the packing parameter of the
  surfactant and thus the tendency of the reverse micelles to solubilize water
  from the droplet. At first sight, we expect the packing parameter to decrease
  in the order of increasing hydrophilic-lipophilic-balance (HLB) value: oleic
  acid (HLB=1), monoolein (HLB=3.8), Span 80 (HLB=4.3). However, this does not
  explain why only monoolein leads to swimming motion.}
\end{thebibliography}

\end{document}